\documentclass[prl,aps,showpacs,twocolumn,superscriptaddress,longbibliography]{revtex4-1}
\usepackage{graphicx}
\usepackage{amssymb}
\usepackage{amsmath}
\usepackage[colorlinks=true,linkcolor=blue,anchorcolor=blue,citecolor=blue,urlcolor=blue]{hyperref}

\begin{document}

\title{Topological and disorder corrections to the transverse Wiedemann-Franz law and Mott relation in kagome magnets}

\author{Xiao-Bin Qiang}
\affiliation{Shenzhen Institute for Quantum Science and Engineering and Department of Physics, Southern University of Science and Technology (SUSTech), Shenzhen 518055, China}
\affiliation{Quantum Science Center of Guangdong-Hong Kong-Macao Greater Bay Area (Guangdong), Shenzhen 518045, China}
\affiliation{Shenzhen Key Laboratory of Quantum Science and Engineering, Shenzhen 518055, China}
\affiliation{International Quantum Academy, Shenzhen 518048, China}

\author{Z. Z. Du}
\affiliation{Shenzhen Institute for Quantum Science and Engineering and Department of Physics, Southern University of Science and Technology (SUSTech), Shenzhen 518055, China}
\affiliation{Shenzhen Key Laboratory of Quantum Science and Engineering, Shenzhen 518055, China}
\affiliation{International Quantum Academy, Shenzhen 518048, China}

\author{Hai-Zhou Lu}
\email{Corresponding author: luhz@sustech.edu.cn}
\affiliation{Shenzhen Institute for Quantum Science and Engineering and Department of Physics, Southern University of Science and Technology (SUSTech), Shenzhen 518055, China}
\affiliation{Quantum Science Center of Guangdong-Hong Kong-Macao Greater Bay Area (Guangdong), Shenzhen 518045, China}
\affiliation{Shenzhen Key Laboratory of Quantum Science and Engineering, Shenzhen 518055, China}
\affiliation{International Quantum Academy, Shenzhen 518048, China}

\author{X. C. Xie}
\affiliation{International Center for Quantum Materials, School of Physics, Peking University, Beijing 100871, China}
\affiliation{Collaborative Innovation Center of Quantum Matter, Beijing 100871, China}
\affiliation{CAS Center for Excellence in Topological Quantum Computation,
University of Chinese Academy of Sciences, Beijing 100190, China}

\date{\today}

\begin{abstract}
The Wiedemann-Franz law and Mott relation are textbook paradigms on the ratios of the thermal and thermoelectric conductivities to electrical conductivity, respectively. Deviations from them usually reveal insights for intriguing phases of matter. The recent topological kagome magnets TbMn$_6$Sn$_6$ and Mn$_3$Ge show confusingly opposite derivations in the Hall measurement. We calculate the topological and disorder corrections to the Wiedemann-Franz law and Mott relation for the Hall responses in topological kagome magnets. The calculation indicates the dominance of the topological correction in the experiments. More importantly, we derive analytic correction formulas, which can universally capture the two opposite experiments with the chemical potential as the only parameter and will be a powerful guidance for future explorations on the magnetic topological matter.
\end{abstract}

\maketitle

\begin{figure}[ht]
\centering
\includegraphics[width=0.44\textwidth]{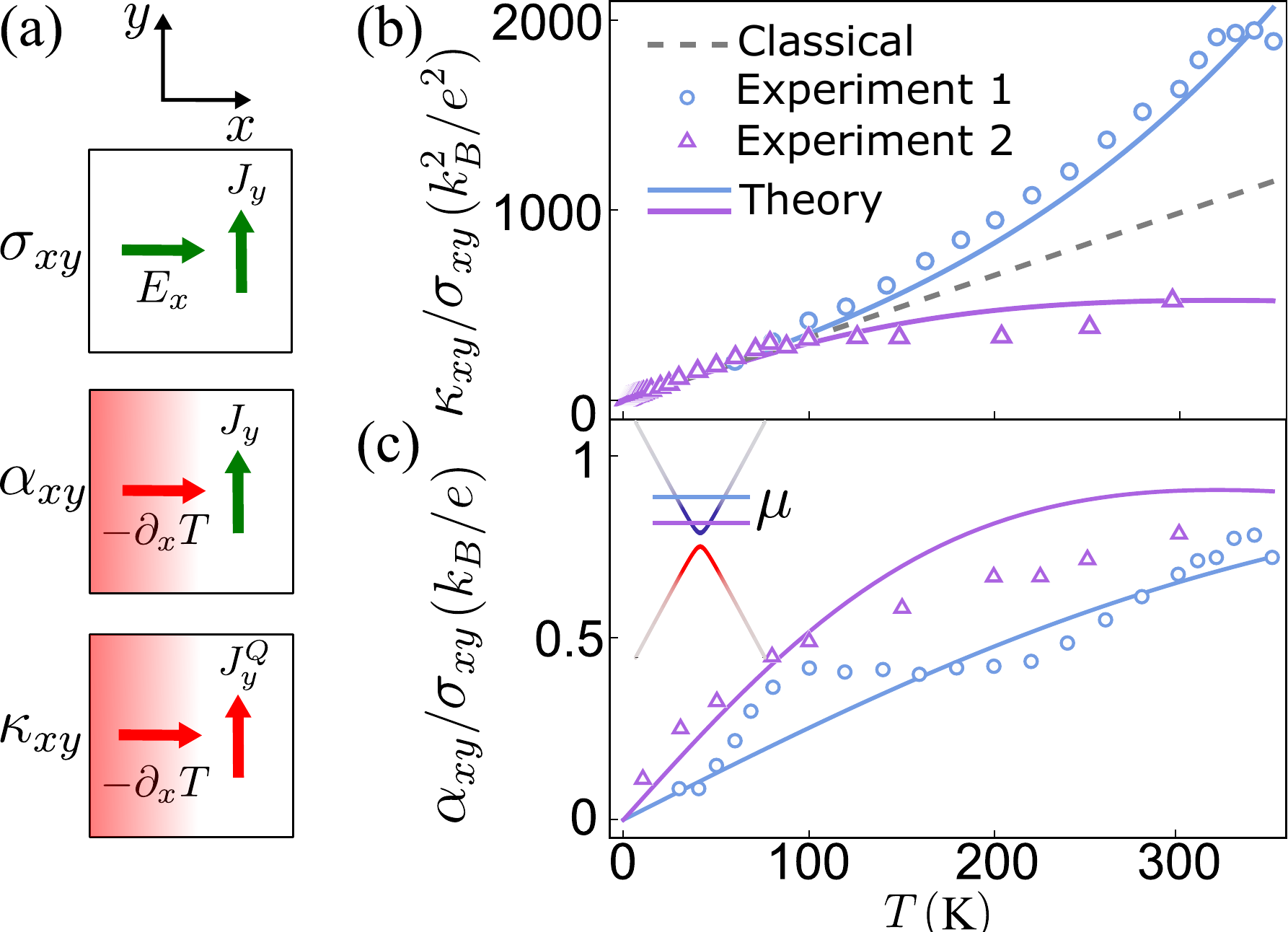} \caption{(a) The electric Hall conductivity $\sigma_{xy}\equiv J_y/E_x$, thermoelectric Hall coefficient $\alpha_{xy}\equiv J_y/(-\partial_x T)$, and thermal Hall conductivity $\kappa_{xy}\equiv J_y^Q/(-\partial_x T)$, as the response of the transverse electric current $J_y$ or thermal current $J_y^Q$ to the longitudinal electric field $E_x$ or temperature gradient $-\partial_x T$. Our analytical corrections (solid lines) to the classical behaviors (dashed lines) of (b) the ratios $\kappa_{xy}^{in}/\sigma_{xy}^{in}$ and (c) $\alpha_{xy}^{in}/\sigma_{xy}^{in}$ can capture two types of experiments (scatters), with the chemical potential $\mu$ as the only tuning parameter. 
We choose $\mu$ = 0.11 and 0.05 eV for the TbMn$_6$Sn$_6$ \cite{Yin22nc} (Exp. 1) and Mn$_3$Ge \cite{Xu20sa} (Exp. 2) and  experiments, respectively. In the TbMn$_6$Sn$_6$ experiment, $\mu=0.13$ eV.}
\label{Fig:main}
\end{figure}

In most metals, the ratio between the thermal conductivity $\kappa$ and electric conductivity $\sigma$ is characterized by the Wiedemann-Franz law $\kappa/\sigma=L_0T$ \cite{Ashcroft76book}, where $T$ is the temperature and the Lorentz number
\begin{equation}\label{Eq:Lorentz_number}
L_0=\left(\pi k_B/e\right)^2/3
\end{equation}
is composed of the elementary charge $e$ and the Boltzmann constant $k_B$.
The deviations from them usually provide insights for exotic phases of matter, such as in non-Fermi liquids and hydrodynamics \cite{ Sean13prb,Lavasani19prb,Kubala08prl,Principi15prl,XiaoC16prb,Lucas18prb,Kim09prl,Buccheri22prb,WangZQ22prb}.
Recently, the thermoelectric and thermal transports have become promising tools to probe topological phases of matter \cite{Xiao06prl,Bergman10prl,Yokoyama11prb,Cong14prb,Max16nm,Sharma16prb,Noky18prb,Nandy19prb,WangCM19prl,Zeng20arXiv,FuCG20aplM,Minami20prb,Zhang21prb}.
In particular, topological magnets have attracted tremendous interest because of their outstanding electromagnetic conversion abilities for potential device applications \cite{Satoru15nature,Lin18prl,Ye18nature,Yin18nature,Yin19np,Yin20nature}.
The latest experiments on the topological kagome magnets TbMn$_6$Sn$_6$ \cite{Yin22nc} and Mn$_3$Ge \cite{Xu20sa} show confusingly opposite deviations from the Wiedemann-Franz law in the Hall measurements [Fig. \ref{Fig:main} (b)], raising questions on their microscopic mechanisms. In particular, disorder and topology could both play sophisticated roles in the anomalous Hall transport \cite{Nagaosa10rmp}, but their contributions and competition remain unclear in these experiments.

\begin{figure}[ht]
    \centering
    \includegraphics[width=0.45\textwidth]{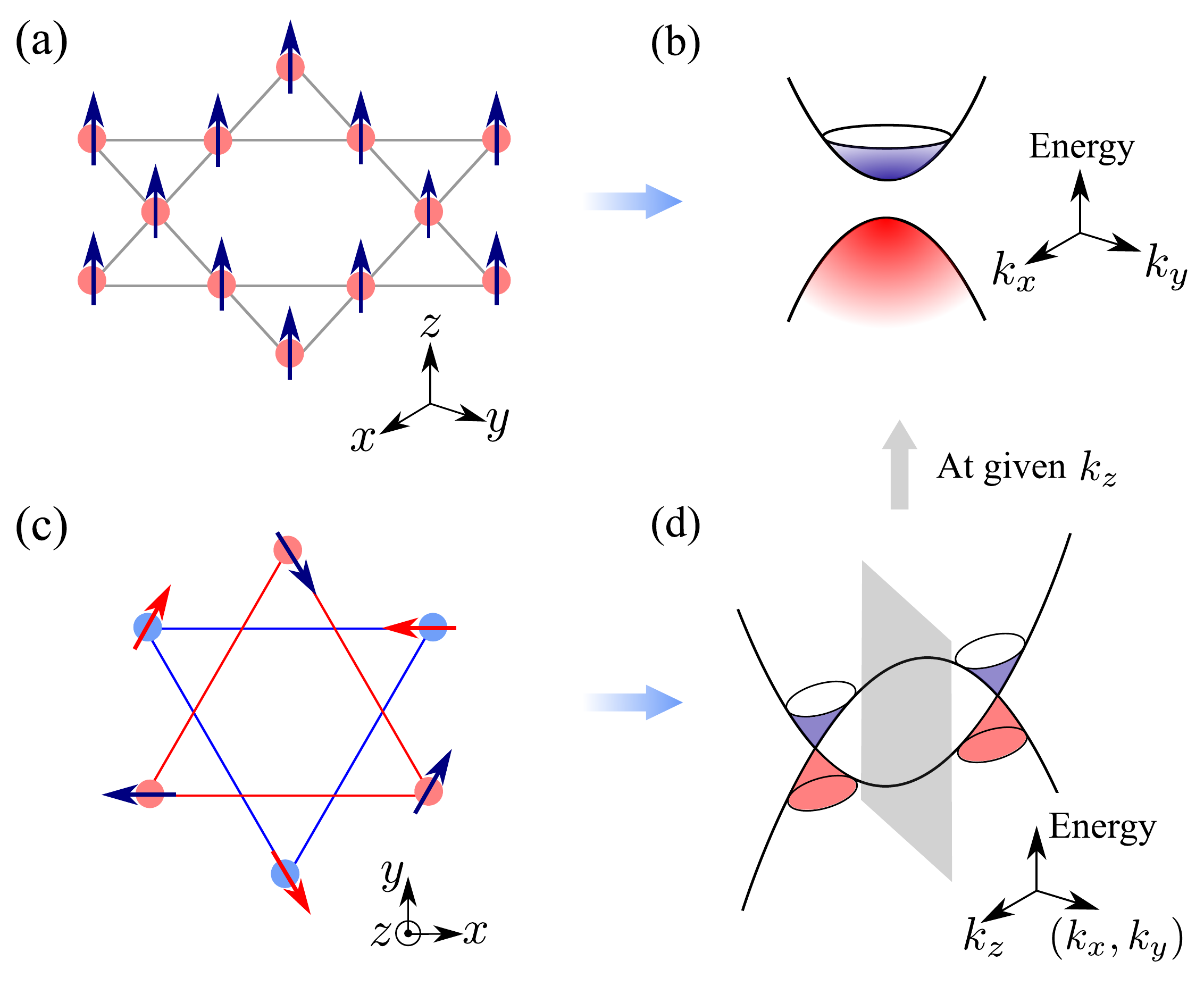} \caption{Two typical topological kagome magnets. (a) TbMn$_6$Sn$_6$ with the out-of-plane magnetization \cite{Yin22nc} can be described by the 2D massive Dirac model (b), as verified by the fan diagram of the Landau levels under magnetic fields \cite{ZhangSC15prl,Yin20nature}. (c) The electronic behaviors of the layered (red for one sublayer, blue for the other) in-plane non-colinear antiferromagnet Mn$_3$Ge \cite{Xu20sa} is
    believed to be described by the Weyl semimetal \cite{kubler14epl}, which can be viewed as layers of coupled 2D massive Dirac models (d).}
\label{Fig:kagome}
\end{figure}

In this Letter, we calculate the ratios
of thermal Hall conductivity $\kappa_{xy}$
and thermoelectric Hall (Nernst) conductivity
$\alpha_{xy}$ to the electric Hall conductivity $\sigma_{xy}$ [Fig. \ref{Fig:main} (a)].
Beyond the previous works, we take into account the disorder contribution and topological contribution on same footing. By comparing the calculation with the experiments, we find that
the topological contribution may dominate in the experiments. More importantly, we can derive the analytic formulas for the corrections to the Wiedemann-Franz law ($\kappa/\sigma$) and Mott relation ($\alpha=-e L_0T \partial\sigma/\partial \mu$, where $\mu$ is the chemical potential), with the help of the Dirac model that carry the topological and magnetic properties of the topological kagome magnets [Fig. \ref{Fig:kagome}].
The analytic formulas  depend not explicitly on the model details, but only on the chemical potential, implying the universal nature of the corrections.
Our analytic formulas can reproduce both the negative and positive derivations from the classical Wiedemann-Franz law [Fig. \ref{Fig:main} (b)], as well as the tendencies in the Mott relation
[Fig. \ref{Fig:main}(c)] in the experiments, with the same chemical potential in each experiment. These formulas will be helpful for further explorations on the electric, thermoelectric, and thermal transports in topological magnets.

Despite the sophisticated magnetic structures of the topological kagome magnets, in the metallic phase as those in the experiments \cite{Yin22nc,Xu20sa}, the magnon \cite{Nagaosa10science,Nagaosa12prb,Neumann22prl} and phonon \cite{Ashcroft76book,Ziman01book,ZhangLF09prb,ZhangLF10prl,Sun21prb} degrees of freedom are inactive in the thermal transport. Therefore, the electronic, thermoelectric, and thermal transports are fairly described by the electronic Hamiltonian. The electronic structure of TbMn$_6$Sn$_6$ can be described by the 2D massive Dirac model, which has been confirmed by the spectroscopic measurement \cite{Yin20nature}. The electronic structure of non-colinear antiferromagnet Mn$_3$Ge [Fig. \ref{Fig:kagome}(c)] \cite{Xu20sa} is the Weyl semimetal \cite{kubler14epl}, which can be viewed as layers of coupled 2D massive Dirac models \cite{Burkov11prl}. More importantly, the massive Dirac model can describe two of the most important characteristics of the topological kagome magnets, i.e., time-reversal symmetry breaking and topological properties. More importantly, later we will see that $\kappa_{xy}^{in}/\sigma_{xy}^{in}$ and $\alpha_{xy}^{in}/\sigma_{xy}^{in}$ in Eqs. (\ref{Eq:WF}) and (\ref{Eq:Mott}) do not depend on the model details.

The electric, thermoelectric (Nernst coefficient), and thermal Hall conductivities can
be expressed in an organized form
\begin{equation}\label{Eq:coes}
\sigma_{xy}^\chi=-\frac{e^2}{h} \mathcal{C}_0^\chi,\ \
\alpha_{xy}^\chi=\frac{k_B e}{h}\mathcal{C}_1^\chi,\ \
\kappa_{xy}^\chi=-\frac{k_B^2 T}{h} \mathcal{C}_2^\chi,
\end{equation}
where $h$ is the Planck constant and the coefficient $\mathcal{C}_n^\chi$ (see \cite{Supp} for details) can be expressed as
\begin{equation}\label{Eq:cn}
\mathcal{C}_n^{\chi}=\int d\varepsilon[\chi]\left(\frac{\varepsilon-\mu}{k_B T}\right)^n\left(-\frac{\partial f^0}{\partial \varepsilon}\right),
\end{equation}
here $\varepsilon$ is the energy, $\mu$ is the chemical potential, and $f^0=1/(e^{(\varepsilon-\mu)/k_BT}+1)$ is the Fermi distribution function. The kernel function $\chi$ denotes different mechanisms in the anomalous transport, including the topological contribution revealed in the seminal works \cite{Xiao06prl,Bergman10prl,Yokoyama11prb} and disorder contribution \cite{Sinitsyn06prl,Sinitsyn07prb,FuL21prb}. For the topological contribution, $\chi^{in}=2\pi\sum_l \Omega_l^zf_l^0$, where $l=(\mathbf{k},n)$ stands for the quantum numbers (momentum, band index), $\Omega^z_l$ is the $z$ component of the Berry curvature of band $n$ with the definition $\Omega_l= i\nabla_\mathbf{k}\times\langle \psi_n|\nabla_{\mathbf{k}} \psi_n\rangle$.
The Berry curvature can be understood as the ``magnetic field" in the parameter space, as a result of the geometric structure of the quantum states \cite{Berry84}, and is found critical in understanding the anomalous Hall effect \cite{Xiao10rmp,Nagaosa10rmp}, topological magnetoresistance \cite{Son13prb,Burkov14prl,Pan15arxiv,Dai17prl}, and thermal transport \cite{Xiao06prl,Bergman10prl,Yokoyama11prb}.
For the 2D Dirac model $\mathcal{H}= v(k_x\sigma_x+k_y\sigma_y)+m\sigma_z$, $\Omega_{\pm}^z=\mp mv^2/(2\varepsilon_+^3)$, where the band dispersions are $\varepsilon_{\pm}(k)=\pm[ v^2(k_x^2+k_y^2)+m^2]^{1/2}$, $ v$ is the model parameter, $\sigma_{x,y,z}$ are Pauli matrices, and $m$ is the band gap.

The main results of this work are the analytic formulas for the topological contributions $\kappa_{xy}^{in}/\sigma_{xy}^{in}$ (the Wiedemann-Franz law of the Hall signals) and $\alpha_{xy}^{in}/\sigma_{xy}^{in}$, which can reproduce two opposite types of experiments of the topological kagome magnets. By performing the Sommerfeld expansion of the transport coefficients (Eqs. \ref{Eq:coes}) with respect to the energy scale of temperature (up to $\sigma\propto (k_BT)^4$, $\alpha\propto (k_BT)^4$, $\kappa\propto (k_BT)^4$), we obtain the analytic expression for the ratio $\kappa_{xy}^{in}/\sigma_{xy}^{in}$
\begin{eqnarray}\label{Eq:WF}
\frac{\kappa_{xy}^{in}}{\sigma_{xy}^{in}}
&=&\frac{15\mu^2/\pi^2+21(k_BT)^2}{15\mu^2/\pi^2+5(k_BT)^2+7\pi^2(k_BT)^4/\mu^2}L_0 T,
\end{eqnarray}
where the Lorentz number $L_0=\left(\pi k_B/e\right)^2/3$, the chemical potential $\mu$ is measured from the center of the gap. This formula can be simplified in two limits \cite{Supp}
\begin{equation}\label{Eq:WF-simplified}
\begin{aligned}
\frac{\kappa_{xy}^{in}}{\sigma_{xy}^{in}}=\left\{\begin{aligned}
    &\left(1+\frac{16\pi^2}{15}\frac{k_B^2T^2}{\mu^2}\right)L_0T,  \qquad \mu\gg \mu_c;\\
    &\left(1+\frac{\pi^2}{3}\frac{k_B^2T^2}{\mu^2} \right)^{-1}L_0T, \qquad \mu \ll \mu_c,
   \end{aligned}
   \right.
\end{aligned}
\end{equation}
where the critical chemical potential is found as $\mu_c=\sqrt{7}\pi k_BT/4$ (at which $\kappa_{xy}^{in}/\sigma_{xy}^{in}$ recovers the classical value $L_0T$). The $\mu \gg \mu_c$ limit of Eq. (\ref{Eq:WF-simplified}) has been used to fit the positive deviation in the experiment of TbMn$_6$Sn$_6$ \cite{Yin22nc}, but only the new formula Eq. (\ref{Eq:WF}) can give both the positive and negative deviations in Fig. \ref{Fig:main}.

Similar to the classical Mott relation \cite{Ashcroft76book}, we also get the relation between the thermoelectric and the electric Hall coefficients
\begin{equation}\label{Eq:Mott}
\frac{\alpha_{xy}^{in}}{\sigma_{xy}^{in}}=\left(\frac{\mu}{e}+\frac{\pi^2}{3}\frac{k_B^2 T^2}{e\mu} \right)^{-1}L_0T.
\end{equation}
Note that the formulas are independent of the model parameters but solely depend on $\mu$, i.e. the position of the chemical potential.
This is reasonable because the model details are cancelled as the quantities on the denominator and numerator of the ratios both depends on the model parameters, such as the band gap $m$. In this sense, the ratios are more intrinsic and universal.

To test Eqs. (\ref{Eq:WF}) and (\ref{Eq:Mott}), we use $\mu$ as the only parameter to fit the experimental data from Refs. \cite{Yin22nc,Xu20sa}.
Figs. \ref{Fig:main} (b) and (c) show a good agreement between the experiments and our analytic formulas for $\kappa_{xy}^{in}/\sigma_{xy}^{in}$ and $\alpha_{xy}^{in}/\sigma_{xy}^{in}$ as functions of temperature $T$. In particular, we use $\mu=0.11$ eV to fit the data of TbMn$_6$Sn$_6$, very close to the measured $\mu=0.13\pm 0.004$ eV in the experiment \cite{Yin20nature}. We stress that the fitting parameter $\mu$ is the same for $\kappa_{xy}^{in}/\sigma_{xy}^{in}$ and $\alpha_{xy}^{in}/\sigma_{xy}^{in}$ for the same experiment in the comparison. Constrained by the Sommerfeld expansion, our formulas in Eqs. (\ref{Eq:WF}) and (\ref{Eq:Mott}) are valid for temperatures below the Fermi temperature $T_F=\mu/k_B $, which is more than 1000 K for the high chemical potentials in the experiments. Nevertheless, there is a derivation from the formula above 300 K.

It is reasonable to employ the formulas, because the electronic behaviors of the topological kagome magnet TbMn$_6$Sn$_6$ \cite{ZhangSC15prl,Yin20nature,Yin22nc}
and the non-colinear antiferromagnet Mn$_3$Ge \cite{kubler14epl,Xu20sa} can be reduced to the 2D massive Dirac model [Fig. \ref{Fig:kagome}].
The description of Mn$_3$Ge by the 2D model is further supported by considering a 3D tilted massless Weyl cone \cite{MaDa19prb,Burkov22prb} $\mathcal{H}=s(v \mathbf{k}\cdot\sigma+tk_z)$, where $v$ is the model parameter, and $s=\pm 1$ labels the chirality of a single node. The tilt term $tk_z$ is necessary to cover more general case, and we assume $t/v\ll 1$ to get the analytical result. This model can well describe the Weyl semimetal phase in the bulk Mn$_3$Ge \cite{Burkov22prb}. We find that there is no deviation from the Wiedemann-Franz law and Mott relation when both the topology and disorder contributions are taken into account for the tilted massless Weyl cone (Details can be found in \cite{Supp}), additionally justifying that the 2D model is a better description for the layered structure of Mn$_3$Ge.


\begin{figure}[htbp]
\centering
\includegraphics[width=0.45\textwidth]{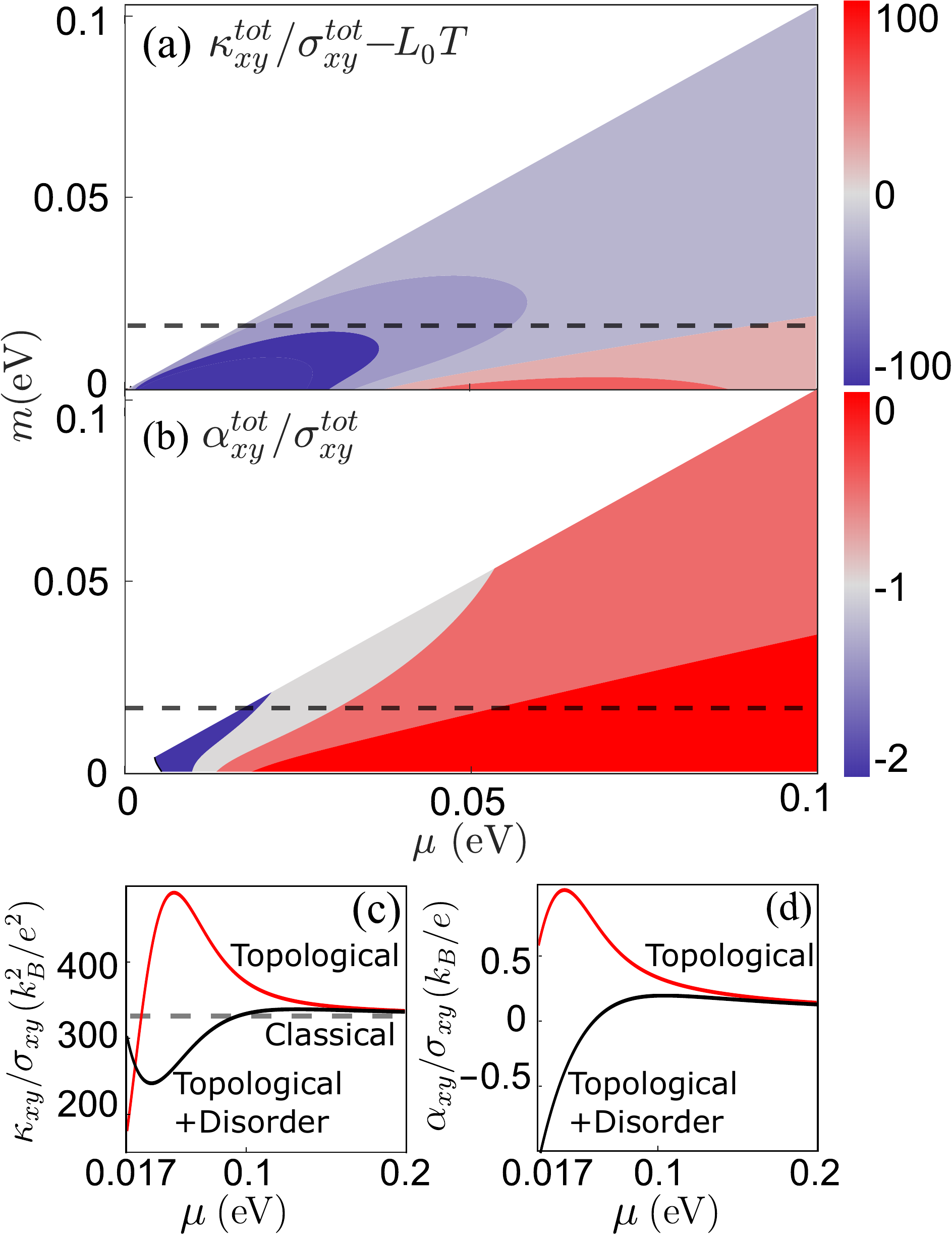}
    \caption{(a) The total (topological $+$ disorder) contribution to the deviation of Wiedemann-Franz law $\kappa_{xy}^{tot}/\sigma_{xy}^{tot}-L_0T$ in unit of $k_B^2/e^2$, and (b) $\alpha_{xy}^{tot}/\sigma_{xy}^{tot}$ in unit of $k_B/e$ for the 2D massive Dirac model. The dashed lines represent $m=0.017$ eV for the TbMn$_6$Sn$_6$ \cite{Yin22nc}. [(c) and (d)] The chemical potential $\mu$ dependence of two ratios for the topological (red) and total (black) contribution with $m=0.017$ eV. The parameters are $ v=1$ eV$\cdot$nm, $n_i V_0^4/V_1^3=1$ eV, and $T=100$ K.}
\label{Fig:Total}
\end{figure}

Up to now, we only consider the topological correction. However, disorder in real materials can contribute to the anomalous transport \cite{Nagaosa10rmp,Sinitsyn06prl,Sinitsyn07prb,Lu13prb,FuL21prb}. We use the Boltzmann kinetics to account for the disorder contribution, which is widely used in explaining magneto-transport \cite{Yip15arXiv,Spivak16prb,Dai17prl,Spivak18prl} and nonlinear transport \cite{Fu15prl,Du18prl,Du19nc,Du21nrp,Su22prb}. To have analytic results, we consider a $\delta$-correlated disorder model $\hat{V}_{imp}=\sum_i V_i \delta(\mathbf{r}-\mathbf{R}_i)$ with the randomly distributed vector $R_i$ and the disorder strength $V_i$ satisfying the second-order correlation $\langle V_i^2\rangle_{dis}=V_0^2$ and third-order correlation$\langle V_i^3\rangle_{dis}=V_1^3$. This model can well describe the elastic scattering (even the electron-phonon scattering approximately below the Debye temperature \cite{Ashcroft76book}). After a lengthy calculation \cite{Supp}, we find the kernel functions for the disorder contributions, including the side-jump, extrinsic skew-scattering, and intrinsic skew-scattering parts, respectively,
\begin{equation}\label{Eq:leading-disorder}
\begin{aligned}
&\chi^{sj}=\frac{2m(\varepsilon^2-m^2)}{ \varepsilon(\varepsilon^2+3m^2)},\\
&\chi^{sk,1}=\frac{V_1^3}{n_iV_0^4} \frac{m(\varepsilon^2-m^2)^2}{(\varepsilon^2+3m^2)^2},\\
&\chi^{sk,2}=\frac{3m(\varepsilon^2-m^2)^2}{2\varepsilon(\varepsilon^2+3m^2)^2}.
\end{aligned}
\end{equation}
By substituting $\chi$'s into Eq. (\ref{Eq:cn}), we can have the total contribution that includes the topological and disorder contributions for all transport coefficients, as well as the ratios $\kappa_{xy}^{tot}/\sigma_{xy}^{tot}$ and $\alpha_{xy}^{tot}/\sigma_{xy}^{tot}$.

Interestingly, in the presence of the disorder contribution, up to the leading-order ($\sigma\propto (k_BT)^0$, $\alpha\propto (k_BT)^2$, $\kappa\propto (k_BT)^2$) Sommerfeld expansion of Eq. (\ref{Eq:cn}), the ratios can recover
the classical Wiedemann-Franz law and Mott relation, i.e., (check \cite{Supp} for details)

\begin{equation}\label{Eq:leading-disorder-WF-Mott}
\begin{aligned}
\kappa_{xy}^{in(0)}+\kappa_{xy}^{dis(0)}
&=L_0 T \left[\sigma_{xy}^{in(0)}+\sigma_{xy}^{dis(0)}\right],\\
\alpha_{xy}^{in(0)}+\alpha_{xy}^{dis(0)}
&= -e L_0T \frac{\partial}{\partial \mu}\left[\sigma_{xy}^{in(0)}+\sigma_{xy}^{dis(0)}\right],
\end{aligned}
\end{equation}
where the superscript $(0)$ means leading-order and $dis(0)$ includes the leading-order side-jump and skew-scattering contributions.

To fully examine the disorder correction, we numerically evaluate $\kappa_{xy}^{tot}/\sigma_{xy}^{tot}$ and $\alpha_{xy}^{tot}/\sigma_{xy}^{tot}$ by using Eqs. (\ref{Eq:cn}) and (\ref{Eq:leading-disorder}) for different values of $m$ and $\mu$. As shown in [Figs. \ref{Fig:Total} (a)-(b)], both $\kappa_{xy}^{tot}/\sigma_{xy}^{tot}$ and $\alpha_{xy}^{tot}/\sigma_{xy}^{tot}$ show the $m$ dependence, which is different when there is only the topological contribution. We adopt the value $m=0.017$ eV for the TbMn$_6$Sn$_6$ \cite{Yin22nc}, i.e., black curves in [Figs. \ref{Fig:Total} (c)-(d)]. It can be seen that the topological and total contribution show different behaviors. For $\kappa_{xy}/\sigma_{xy}$, the total contribution show almost only negative correction to the Wiedemann-Franz law, unable to describe the positive deviation in the TbMn$_6$Sn$_6$ experiment \cite{Yin22nc}. Similarly, if we include the disorder contribution, the total correction to the
$\alpha_{xy}/\sigma_{xy}$ is also not consistent with the experiment \cite{Yin22nc}. These results imply that the topological contribution is dominant in these topological kagome magnet experiments so far \cite{Yin22nc,Xu20sa}.

\begin{figure}[htbp]
\centering
\includegraphics[width=0.45\textwidth]{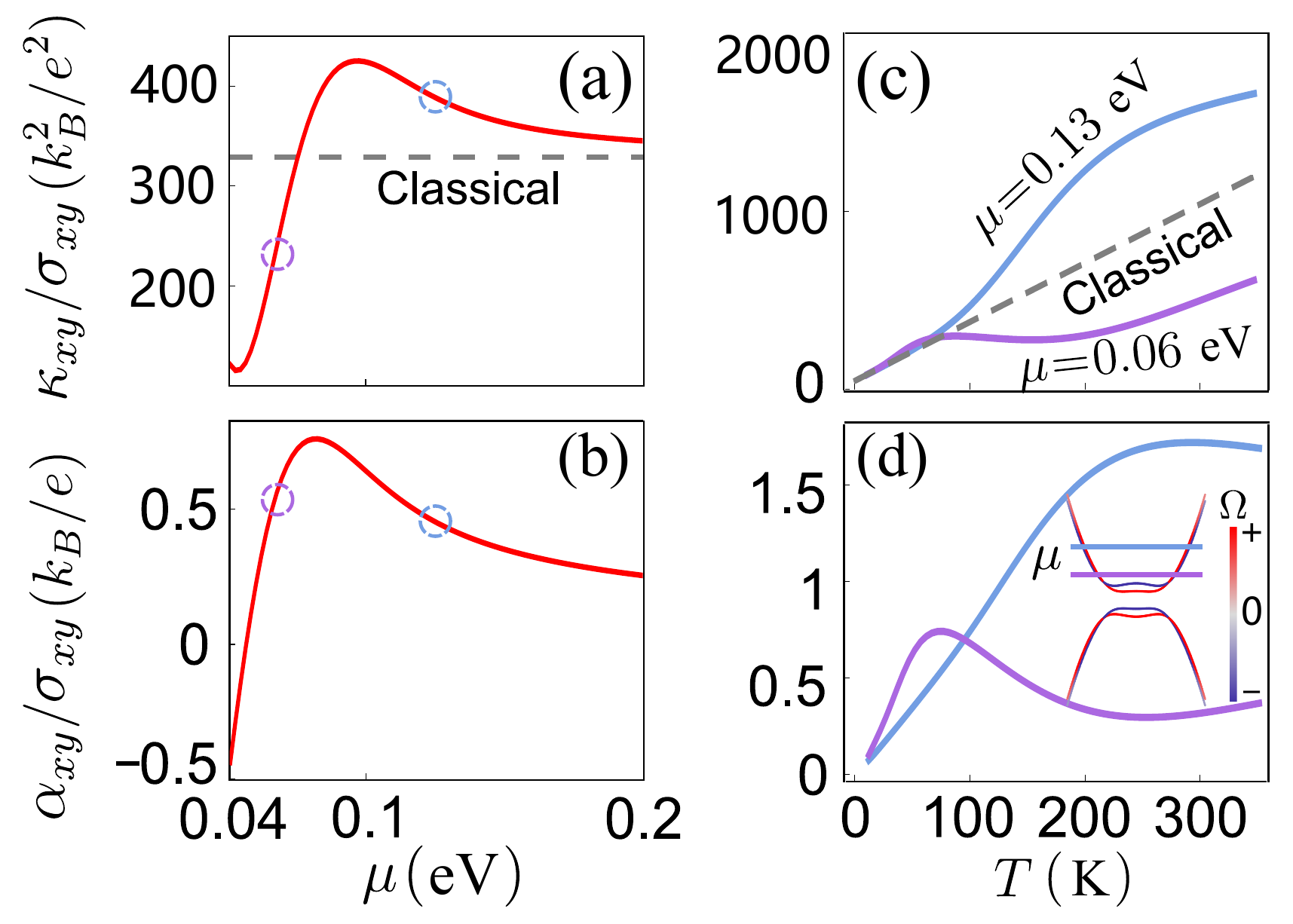}
    \caption{[(a) and (b)] For the 3D Dirac model [Eq. (\ref{Eq:TI})], $\kappa_{xy}^{in}/\sigma_{xy}^{in}$ (a) and $\alpha_{xy}^{in}/\sigma_{xy}^{in}$ (b) as functions of the chemical potential $\mu$ at 100 K. The dashed lines represent the classical Wiedemann-Franz law. The purple and blue circles indicate $\mu=0.06$ eV and $\mu=0.13$ eV, corresponding to the two curves in (c) and (d). [(c) and (d)] $\kappa_{xy}^{in}/\sigma_{xy}^{in}$ (c) and $\alpha_{xy}/\sigma_{xy}$ (d) as functions of temperature $T$, for $\mu=0.06$ eV and $\mu=0.13$ eV (the solid curves). Inset of (d) depicts the dispersion and Berry curvature $\Omega$ of the energy bands for the 3D Dirac model with the Zeeman energy, which lifts the degeneracy of the bands. The parameters are $v=0.1$ eV$\cdot$nm, $m=-0.04$ eV, $b=-0.18$ eV$\cdot$nm$^2$, and $M=5.788\times10^{-3}$ eV \cite{Fu19prl}.}
\label{Fig:TI}
\end{figure}


To see if the above results based on the 2D Dirac model can be generalized to 3D, we also use a 3D Dirac model to calculate the topological correction to $\kappa_{xy}^{in}/\sigma_{xy}^{in}$ and $\alpha_{xy}^{in}/\sigma_{xy}^{in}$ \cite{Shen17book}
\begin{equation}\label{Eq:TI}
\mathcal{H}=v \mathbf{k}\cdot\boldsymbol{\alpha}+(m-bk^2)\beta+M\sigma_0\otimes \sigma_z,
\end{equation}
where $v, m$, $b$, and $M$ are model parameters, $\alpha_i=\sigma_x\otimes\sigma_i$ and $\beta=\sigma_z\otimes\sigma_0$ are Dirac matrices, $\mathbf{k}=(k_x,k_y,k_z)$ is the wavevector with $k=|\mathbf{k}|$.
When the parameters $mb>0$, this model can characterize a magnetic topological insulator. The last term is the Zeeman energy that describes magnetism, which
breaks time-reversal symmetry to unveil the hidden Berry curvature, as seen in inset of Fig. \ref{Fig:TI} (d).

In Fig. \ref{Fig:TI}, we present the numerical results of $\kappa_{xy}^{in}/\sigma_{xy}^{in}$ and $\alpha_{xy}^{in}/\sigma_{xy}^{in}$ for the 3D Dirac model in Eq. (\ref{Eq:TI}), which show similar temperature and chemical potential dependences to those described by the 2D Dirac model and in the experiments. Specifically, the violation of the classical Wiedemann-Franz law becomes stronger as the chemical potential approaches the band edge and as temperature is increased, and $\kappa_{xy}^{in}/\sigma_{xy}^{in}$ is below and above the classical value, respectively, as the chemical potential is very close to and moves away from the band edge.
These numerical results of 3D Dirac model show that our analytical formulas based on the 2D Dirac model could be a qualitative tool to study the topological corrections to the Wiedemann-Franz law and Mott relation in the topological magnets.

To conclude, we analytically and numerically calculated the corrections to the  Wiedemann-Franz law and Mott relation for topological kagome magnets, by treating the topological correction and disorder correction on same footing. By comparing our results with the recent experiments with opposite behaviors, we show the dominance of the topological correction from the Berry curvature in the experiments. More importantly, our analytic formulas for the topological correction will be an useful tool to explore the emergent topological magnets.

We thank Jia-Xin Yin for helpful discussions. This work was supported by the National
Key R$\&$D Program of China (2020YFA0308900), 
Innovation Program for Quantum Science and Technology (2021ZD0302400),
the National Natural Science Foundation of China (11925402),
the National Basic Research Program of China (2015CB921102), the Strategic Priority Research Program of Chinese Academy of Sciences (XDB28000000),
Guangdong province (2020KCXTD001 and 2016ZT06D348), 
and the Science, Technology and Innovation Commission of Shenzhen Municipality (ZDSYS20170303165926217, JAY20170412152620376, and KYTDPT20181011104202253). The numerical calculations were supported by Center for Computational Science and Engineering of SUSTech.

\bibliography{ref}

\end{document}